\documentclass[twoside,twocolumn]{article}

\usepackage{abstract} 
\usepackage{amsmath} 
\usepackage[english]{babel} 
\usepackage{booktabs}
\usepackage[hang, small,labelfont=bf,up,textfont=it,up]{caption} 
\usepackage{enumitem} 
\usepackage{fancyhdr}
\usepackage[T1]{fontenc}
\usepackage[hmarginratio=1:1,top=32mm,columnsep=20pt]{geometry} 
\usepackage{hyperref} 
\usepackage[sc]{mathpazo} 
\usepackage{microtype} 
\usepackage{titlesec} 
\usepackage{titling}
\usepackage{url} 

\renewcommand\thesection{\Roman{section}} 
\renewcommand\thesubsection{\roman{subsection}} 

\setlist[itemize]{noitemsep} 

\titleformat{\section}[block]{\large\scshape\centering}{\thesection.}{1em}{} 
\titleformat{\subsection}[block]{\large}{\thesubsection.}{1em}{} 

\title{\huge Challenges of Proof-of-Useful-Work (PoUW)} 
\author{%
\textsc{Felix Hoffmann} \\[1ex] 
\normalsize Johann Wolfgang Goethe-Universität Frankfurt am Main \\ 
\normalsize \href{mailto:felix.hoffmann@iri.uni-frankfurt.de}{felix.hoffmann@iri.uni-frankfurt.de} 
}
\date{\today} 
\renewcommand{\maketitlehookd}{
\begin{abstract}
\noindent
Proof-of-Work (PoW) is a popular blockchain consensus algorithm that is used in cryptocurrencies like Bitcoin in which hashing operations are repeated until the resulting hash has certain properties. This approach uses lots of computational power and energy for the sole purpose of securing the blockchain. In order to not waste energy on hashing operations that do not have any other purpose than enabling consensus between nodes and therefore securing the blockchain, Proof-of-Useful-Work (PoUW) is an alternative approach which aims to replace excessive usage of hash functions with tasks that bring additional real-world benefit, e.g. supporting scientific experiments that rely on computationally heavy simulations. This publication consists of two parts: In the first part, important properties of conventional hash-based PoW are described. In the second part, theoretical PoUW concepts such as Coinami, CoinAI and the first successful PoUW cryptocurrency Primecoin are analyzed with respects to how PoW properties can be retained while doing useful work.
\end{abstract}
}

\begin{document}
\maketitle

\section{Introduction}
Traditional proof-of-work cryptocurrencies have been widely criticized for using up lots of energy in order to run and secure the underlying blockchain. In the past few years, there has been done research with the goal of replacing repeated hash operations with useful work. Notable projects such as Primecoin \cite{primecoin_whitepaper}, Coinami \cite{coinami_whitepaper} and CoinAI \cite{coinai_whitepaper} use search for certain kinds of prime number chains, multiple sequence alignment of protein sequences or training of deep learning models as useful work consensus algorithms. In order to give an overview of the challenges these projects have to overcome, the following part gives an outline of important properties that hash-based PoW solutions have.


\section{Properties of hash-based PoW}
Hash-based PoW consensus algorithms use cryptographically secure hash functions such as SHA256 or KECCAK256 in order generate hex strings of fixed size. The proof-of-work hash puzzle consists of finding a result that is smaller than a given number which defines the difficulty of the problem. Hash functions are one-way functions which also allow for quick verification of a proposed solution's validity. If a hash function is not considered to be broken, there is no known way to manipulate the hash function's input to influence its output in a preferred direction. Therefore, the nodes of the blockchain have to brute-force different inputs until they solve the hash puzzle by luck. This leads to an arms race between miners, in which more hardware is acquired to increase ones hash rate. For non-ASIC-resistant cryptocurrencies such as Bitcoin, specialized hardware with the sole purpose of efficiently mining can be used.\\In the following, advantageous properties of hash-based PoW algorithms are outlined and it is described why these properties are useful in the context of blockchains.

\subsection{Block sensitivity \& non re-usability}
In order to prevent the re-usage of existing proofs-of-work, it is necessary to bind the validity of a PoW to the block it validates. This means that the computational work has to factor in information that is not known beforehand to prevent pre-calculation of future blocks. A common strategy to retain the block sensitivity property is to require the hash of the previous block to be used as part of the hash function's input for the PoW of the next block. Since the hash of the previous block is only known when it is successfully added to the blockchain, this makes it infeasible to pre-calculate future blocks as long as the hash function used is not fundamentally broken. However, there exists an attack called \textit{Selfish Mining} in which a group of malicious miners that solved the current hash puzzle do not broadcast their solution to all miners but instead continue mining additional blocks in secret until they decide to publish a long chain of new blocks. Since common blockchain implementations obey the \textit{longest chain rule}, all other valid blocks that were mined by other nodes during this time will be discarded if the public state of the chain is shorter than the secretly mined chain. Selfish Miners always run the risk of not being able to keep up with the speed of the chain they compete against, in which case block rewards that could have been collected by Selfish Miners are lost. As a result, Selfish Mining strategies are only feasible if large amounts of a blockchain network's computational power is controlled by an organized group of miners.

\subsection{Adjustable problem hardness}
The difficulty of the PoW needs to be adjustable. This property is required so that block intervals can be regulated (e.g. Bitcoin's average block time is around 10 minutes \cite{bitcoin_whitepaper}). The goal is to both counteract inflation\footnote{Additionally, in the case of Bitcoin, block rewards are decreased over time to further prevent inflation.} (miners of blocks are financially compensated for their work) and to guarantee a stable transaction throughput. Problem hardness commonly is dynamically adjusted over time depending on the current total hash rate of the network. It should be noted that the difficulty of PoW problems need to have a lower bound: Trivial problems that can be solved instantly are not suitable for a consensus algorithm because then miners are incentivized to mine empty blocks instead of filling them with pending transactions. Using hash-based PoW approaches has the advantage that the difficulty of problems can trivially be adjusted in both directions by lowering/increasing the required upper bound of accepted hash values.\\\\
Another property of hash-based PoW is that miners with limited computational power have a non-zero chance of quickly solving the hash puzzle by luck. If an entity with computational power $\alpha$ would always lose against entities with computational power $\beta > \alpha$, then the blockchain would be dictated by the single largest group of miners, which as a result would disincentivize miners from participating in the blockchain. This would be the death of the blockchain since known attacks like the 51\% attack would become feasible. Therefore, it is not only important that in the long term a miner earns block rewards that are proportional to that miners hash rate contribution to the overall network, but also that any miner has a non-zero chance to successfully solve the proof-of-work puzzle.

\subsection{Fast verification}
In order to be able to find consensus and be protected from spam attacks, miners need to be able to quickly verify the validity of blocks proposed by other miners. Therefore, it is crucial that the proof-of-work can efficiently be verified in reasonable time without demanding excessive computational resources. The need for fast verification mechanisms is the main factor why hash functions are commonly used in proof-of-work algorithms. Executing one SHA256 or KECCAK256 function call on a small input barely uses any computational power, since the main difficulty is finding an input for such a function that produces the required output. Thus, the term \textit{one-way function}.

\subsection{Problem is parallelizable}
In order to make efficient use of existing hardware, it is preferred to use proof-of-work problems that can be parallelized. For instance, finding hash function outputs that have a certain of amount of leading zeroes is called an \textit{embarrassingly parallel problem} since there is no need for communication between threads.\\Further, parallelizable problems enable the formation of mining pools: Depending on the difficulty of the hash puzzle, low hash rate miners might have a probability close to zero to mine a new block alone. By joining existing mining pools in which computational power of multiple entities is combined and block rewards are shared proportionally to the provided hash rate of every pool member, weak miners can collect small amounts of financial compensation in regular intervals.\\ All in all, while a proof-of-work consensus algorithm does not necessarily have to be parallelizable, this property makes mining more accessible for a wider range of participants which positively affects network diversity and strengthens the blockchain's overall security.


\section{Proof-of-Useful-Work (PoUW)}
This section consists of two parts: In the first part, existing PoUW approaches and ideas are briefly introduced. In the second part, they are analyzed with regards to how the properties of hash-based PoW consensus algorithms are retained and which issues might occur.\\Even though exotic consensus algorithm classes like Proof-of-Storage can be considered useful, the focus in this publication is on computationally-heavy PoUW which shares lots of similarities with hash-based PoW.

\subsection{Primecoin}
Primecoin is a PoUW cryptocurrency that was launched in 2013 by Sunny King.\cite{primecoin_whitepaper} \\Its PoUW consists of finding certain types of prime number chains, so-called \textit{Cunningham} and \textit{bi-twin} chains. Cunningham chains are a series of prime numbers that nearly double each time. In mathematical terms, a prime chain of length $n \in \mathbb{N}$ must fulfill 
\begin{equation} \label{fist_order_cunningham}
p_{i+1} = 2p_i + 1
\end{equation}
to be considered a first order chain or 
\begin{equation} \label{second_order_cunningham}
p_{i+1} = 2p_i - 1
\end{equation}
to be considered a second order chain for all $1 \leq i < n$. For instance, $\{41, 83, 167\}$ is a first order chain of length $n=3$ and $\{7, 13\}$ is a second order chain of length $n=2$.\\In addition to Cunningham chains, the third type of chain that Primecoin allows as proof-of-work are bi-twin chains. These are prime chains that consist of a strict combination of first and second order Cunningham primes. The mathematical definition of a bi-twin chain of length $k+1$ is the sequence $\{ n-1, n+1, 2n - 1, 2n + 1, 2^{2}n - 1, 2^{2}n + 1, ... , 2^{k}n - 1, 2^{k}n + 1 \}$.\\For instance, choosing $n=6$ leads to $\{ 5, 7, 11, 13 \}$ which is a bi-twin chain of length 2 that consists of 4 prime numbers.\\\\As of writing this publication, a Primecoin is traded for about \$0.04 and the currency's total market capitalization is around \$1.7 million. \cite{primecoin_coinmarketcap} The success of Primecoin can be seen as evidence that PoUW is a viable concept with real-world applications. 
\subsection{Coinami}
In 2016, a theoretical proposal of a mediator interface for a volunteer grid similar to BOINC middleware that can be connected to a cryptocurrency was published and named Coinami. \cite{coinami_whitepaper} The PoUW of Coinami is built on DNA sequence alignment (HTS read mapping in particular) and aims to generate and analyze huge datasets of disease signatures which can help us to gain a better understanding of diseases such as different cancer variants.\\The authors of Coinami describe their approach as a three-level multi-centric system which consists of a root authority, sub-authorities and miners. Miners download problem sets from sub-authorities, map HTS reads to a reference genome and send the results back to sub-authorities for verification. Sub-authorities are certified by the root authority. \cite{coinami_whitepaper}\\\\As a result, this approach can be seen as a hybrid of Proof-of-Authority (PoA) and Proof-of-Useful-Work (PoUW) consensus algorithms. As of writing, while Coinami does have a prototype implementation on Github \cite{coinami_github}, there currently exists no cryptocurrency that is connected to this academic proposal. 

\subsection{CoinAI}
In 2019, a theoretical proposal of PoUW consensus that is built on training and hyperparameter optimization of deep learning models was published and named CoinAI. \cite{coinai_whitepaper}\\
The goal of CoinAI is to secure a blockchain-based cryptocurrency with a consensus algorithm that both secures the underlying blockchain while also producing deep learning models that solve real-world problems. The proposed proof-of-work consists of training a model that passes a certain performance threshold in order for it to be considered valid. In addition to the training of deep learning models, the CoinAI proposal features another financial incentive to participate in the blockchain: Nodes can rent out available hard drive storage to provide distributed storage for the resulting deep learning models of the blockchain. \cite{coinai_whitepaper}\\\\
Thus, CoinAI's approach can be described as a hybrid of Proof-of-Useful-Work (PoUW) and Proof-of-Storage (PoS). As of writing, CoinAI remains an academic proposal that has not yet been implemented to secure a tradeable cryptocurrency.

\subsection*{Analysis of PoUW approaches}
\subsubsection*{PoUW: Non re-usability}
To prevent future calculation and re-usability of proofs-of-work, a given problem must involve information or parameters that can not reliably be guessed beforehand. All nodes must be able to agree on how these parameters are to be adjusted over time so that the problem sets are adjusted over time and it can be decided whether a given proof-of-work is valid for some time interval. A common approach here is to involve the hash of the previous block as a parameter as part of the next problem. However, since this directly influences the result of the calculations, it must be decided on a case-per-case basis whether the resulting information can still be considered to be useful.\\If incorporating hashes into the calculations is not possible, then another approach must be found to bind the PoUW to a given period in time. Relying on an external (as in information taken from outside the blockchain) source that continuously publishes new information over time is not desirable, since this approach leads to a high degree of centralization which not only opposes core principles of a decentralized blockchain but which also has the potential to create security issues and conflicts of interest, especially if the underlying blockchain is connected to a cryptocurrency.\\\\
$\triangleright$ Primecoin retains the property of block sensitivity by requiring the origin of the prime chains to be divisible by the hash of the previous block. In this case, the resulting quotient is defined as a so-called \textit{PoW certificate}. \cite{primecoin_whitepaper} This guarantees that pre-calculation of future blocks is not a viable strategy as long as there is no scientific breakthrough in efficiently calculating certain chains of large primes.\\\\
$\triangleright$ The theoretical Coinami approach tries to evade re-usability and pre-calculation problems by relying on an authority approach, in which miners must request tasks from (sub)-authority nodes. Since miners can not guess which task they might be given next, pre-calculation of future blocks is not feasible. Since sub-authorities know which problems have already been given out, re-usability is not an issue either. The main issue of this solution can be seen as a high degree of centralization which forces miners to trust any (sub)-authority.\\\\
$\triangleright$ The CoinAI proposal concatenates information such as previous block hash, a random number called nonce and a list of pending transactions which then is hashed. This hash result then is used to determine the initial hyperparameter structure of a deep learning architecture which must be trained until it satisfies performance requirements. An issue that potentially arises with this approach is that if the goal is to produce useful deep learning models, then starting the training with an inadequate initial hyperparameter configuration affects the amount of training required to reach acceptable model performance which can be seen as wasted energy. Assuming that the space of all allowed hyperparameter configurations is limited to prevent this from happening, the next problem that might arise is that now hash-to-hyperparameter-configuration mapping collisions are bound to happen more frequently, which in this case means that multiple hashes lead to the same initial hyperparameter configuration which as a result could make pre-calculation strategies feasible.

\subsubsection*{PoUW: Adjustable hardness}\label{adjustable_hardness}
Since miners might join or leave the network of nodes at any time, the blockchain's total computational power fluctuates over time. In order to provide regular block intervals which in the case of a cryptocurrency is necessary to stabilize the transaction throughput, there must be consensus between nodes with respect to how the difficulty of problems is to be adjusted over time. Hash-based PoW approaches control the problem difficulty by dynamically adjusting the amount of leading zeroes that the resulting hash must have in order to be valid depending on the current hash rate of the network. Increasing the amount of required leading zeroes by just one increases the difficulty of the hash puzzle exponentially, which is why softer variations of this approach can be used (such as e.g. amount of leading digits smaller than eight) to provide a more fine-grained control of the problem difficulty.\\For useful work approaches, it needs to be decided on a case-per-case basis how the hardness of a given problem can dynamically be adjusted without jeopardizing usefulness of results.\\\\
$\triangleright$ In the context of Primecoin, two intuitive mechanics to control problem difficulty come to mind: First of all, the size of prime numbers that start a chain could be increased over time. However, the prime number theorem states that 
\begin{equation} \label{prime_number_theorem}
\lim\limits_{x\to \infty} \frac{\pi(x)}{\frac{x}{\ln(x)}} = 1
\end{equation}
with $\pi(x)$ being the so-called prime-counting function. The for our context useful interpretation of this equation is that the prime density approaches zero, which means that the proof-of-work difficulty over time might become too high to sustain stable transaction throughput long-term.\\The second intuitive approach that comes to mind is to dynamically adjust the required length of valid prime number chains to control the problem difficulty. This is the approach Primecoin takes: Given a prime chain of some length, Primecoin dynamically adjusts its Fermat primality test which results in a relatively linear continuous difficulty function (as opposed to the non-linear difficulty function of the first approach) that is claimed to be accurate enough to adjust the problem hardness appropriately over time. \cite{primecoin_whitepaper}\\\\
$\triangleright$ The Coinami authors have not yet defined how the difficulty of the DNA sequence alignment problems can be dynamically adjusted over time. The issue here is that the network must rely on an external source for HTS data and simply increasing the size of assignments potentially leads to issues with resulting data size and networking bottlenecks. An idea here is to let miners solve multiple problems at once and then let authority nodes randomly select one of these solutions and discard the others. While this can be seen as a waste of useful work it might be necessary sacrifice to control problem difficulty without increasing data sizes.\\\\
$\triangleright$ CoinAI aims to adjust the PoUW difficulty over time by dynamically adjusting the required performance requirements of resulting deep learning models over time. The idea behind this approach is that validating the performance of a given model is less computationally expensive than training the model. An issue with this approach is that even when knowing the network's total computational power, it would be difficult to estimate an adequate performance threshold. With respect to this problem, Coinami authors note that even slightly increasing the difficulty can potentially result in unsolvable problems. Another problem here is that a centralized entity is supposed to collect all submitted models, test their performance and then announce the winner. A negative aspect here is that miners would be forced to trust a centralized authority. If no such authority were to be involved, then other issues would occur: Deep learning models that solve non-trivial problems can have a size from a few megabytes to many gigabytes. If there were no centralized authority, then every node would be forced to download the models of all other nodes and test the performance of all of them in order to determine the winner model. As a result bandwidth limitations, spam and sybil attacks potentially make this approach infeasible.

\subsubsection*{PoUW: Verification}
A core principle of consensus algorithms in public blockchains is that they are used in order to provide nodes with a method that enables them to form consensus about the current state of the blockchain without having to rely on trust. Hash functions are useful in this regard since the validity of a proposed (input, output) tuple can quickly be verified. As soon as hash-based approaches are discarded in favor of methods that perform useful work, it can become difficult to find a verification method that does not have to rely on a verification-by-replication approach in which the entire useful work process has to be repeated by many nodes. For a given problem there might or might not exist a probabilistic verification approach in which the likelihood of some proposed solution being valid can be estimated efficiently. Therefore, it needs to be decided on a case-per-case basis what is the best way to formulate a PoUW problem in such a way that verification of results can happen quickly and with reasonable amounts of computational effort.\\\\
$\triangleright$ In the case of Primecoin, \textit{probable primality} of prime chains is verified using a combination of both the \textit{Fermat} and the \textit{Euler-Lagrange-Lifchitz} test for prime numbers. These are proven mathematical methods that can be used to efficiently verify the primality of a given number with the downside that there exist so-called \textit{pseudoprimes} that pass those prime tests but which are in fact not prime numbers. The authors of Primecoin have concluded that the probability of pseudoprimes occurring is low enough that this issue can be traded in favor of being able to provide a fast and efficient verification mechanism. \cite{primecoin_whitepaper}\\\\
$\triangleright$ In the Coinami proposal, sub-root authorities collect results from miners and verify the validity of alignments using decoy reads that have been placed into the problem. These decoys are planned to make up around 5\% of each problem and they can be pre-calculated by the sub-authorities. After verification, decoy data is removed from the results. The main challenge here is to place decoy data in such a way that miners are not able to spot these segments in their assignments. If a sub-authority has validated a miners solution, then the data is signed and sent back so that it can be added to the blockchain.\\\\
$\triangleright$ In CoinAI resulting deep learning models are considered to be valid proofs-of-work only if they pass the current performance threshold. The authors provide no concrete plans about whether a centralized entity is responsible for verification or if every miner has to verify all submitted models by other nodes. Potential issues that might occur in either case have already been presented in the adjustable hardness section of this publication. \ref{adjustable_hardness} \\A common approach to validate the performance of a deep learning model is to use two separate datsets, one containing training data used for training the model and the second dataset being the validation/test dataset. CoinAI gives no specifics on how nodes acquire required training datasets which potentially poses a challenge in overcoming issues such as networking bottlenecks due to large datasets that need to be downloaded. The current state-of-the-art in training of deep learning models boils down to the fact that you need more and more training data to improve your model over time, since hyperparameter tuning of a model that was trained on a small dataset alone rarely results in a robust model than can reliably solve non-trivial problems. As a result, the training dataset would have to be extended over time which raises further questions about who provides this data, how this affects centralization and who is willing to sacrifice computational power and network bandwidth to test the performance of all submitted models. Even if all of these potential issues were to be resolved, assuming the same model is trained over many blocks one could argue that as soon as better performing models for a given task are discovered all previously published models lose their usefulness since they perform worse than the newer models. This raises the question if such an approach can be considered to be useful work in the first place. If, however, completely different deep learning models are to be trained at regular block intervals, potential problems of continuously broadcasting new training data sets and generating robust models performance might become overwhelming.

\subsubsection*{PoUW: Parallelizability}
In order to enable the efficient usage of multi-core CPUs, GPUs and facilitate the existence of mining pools, a PoUW consensus algorithm preferable should be of \textit{embarrassingly parallel} nature. An intuitive example of such a problem is any form of processing or generation of unrelated data, like it is done in e.g. brute-force searches.\\There are many non hash-based approaches that fulfill this property: For instance, Monte Carlo event generation and reconstruction in particle physics, pattern matching over DNA sequences in bioinformatics and hyperparameter tuning in deep learning can all be considered to be embarrassingly parallel problems.\\\\
$\triangleright$ In Primecoin the search for prime chains can trivially be implemented in a parallelizable way.\\\\
$\triangleright$ Pattern matching over DNA sequences in bioinformatics like proposed in Coinami is of embarrassingly parallel nature.\\\\
$\triangleright$ Training deep learning models and hyperparameter tuning like proposed in CoinAI is an embarrassingly parallel problem.\\\\
All in all, it can be concluded that retaining the parallelizability property is not an issue for PoUW approaches.


\section{Conclusion}
This publication has provided an overview over essential properties that conventional hash-based proof-of-work consensus algorithms possess. Additionally, an analysis of which measures were taken by existing PoUW approaches such as Primecoin, Coinami and CoinAI  in order to retain hash-based PoW properties while rewarding useful work was provided. It was concluded that domain-specific knowledge is required to make PoUW consensus possible and that implementation details must be decided on a case-by-case basis using domain knowledge from that area of research.\\\\The main weakness that all presented PoUW approaches have in common is the verification of results. While the author of Primecoin was able to find an elegant probabilistic solution of this problem, theoretical publications like Coinami and CoinAI had to make both efficiency and decentralization sacrifices to prevent potential problems.\\\\A common issue with designing new PoUW consensus approaches is that the size of resulting data can be significant compared to hash-based approaches which leads to situations in which data must either be stored externally or on-chain which negatively affects not only storage requirements of full nodes but also sync times of new nodes which effectively raises the entry barriers of participating in the blockchain. \\\\All in all, problems of mathematical nature seem to be best suited for PoUW. These problems have the advantage that a large repertoire of probabilistic verification methods already exists for a wide range of problems, which in addition to a generally asymmetrical ratio of computational effort and size of resulting output make this class of problems potential suitable for making PoUW consensus mainstream.\\\\It remains to be seen whether the concept of Proof-of-Work itself will survive the surge of alternative blockchain consensus algorithms like Proof-of-Stake which do not require notable amounts of computational effort to efficiently form consensus and therefore secure the underlying blockchain.

\bibliographystyle{alpha}
\bibliography{references}

\newcommand{\etalchar}[1]{$^{#1}$}
\begin{thebibliography}{IOG{\etalchar{+}}16}

\bibitem[BS19]{coinai_whitepaper}
Alejandro Baldominos and Yago Saez.
\newblock Coin.ai: A proof-of-useful-work scheme for blockchain-based
  distributed deep learning.
\newblock {\em Entropy}, 21(8):723, 2019.

\bibitem[Coi16]{coinami_github}
Coinami.
\newblock Coinami prototype.
\newblock \url{https://github.com/coinami/coinami-pro}, 03 2016.

\bibitem[Coi22]{primecoin_coinmarketcap}
Coinmarketcap.
\newblock Primecoin.
\newblock \url{https://coinmarketcap.com/currencies/primecoin/}, 08 2022.

\bibitem[IOG{\etalchar{+}}16]{coinami_whitepaper}
Atalay~Mert Ileri, Halil~I. Ozercan, Alper Gundogdu, Ahmet~K. Senol, M.~Yusuf
  Oezkaya, and Can Alkan.
\newblock Coinami: A cryptocurrency with dna sequence alignment as
  proof-of-work.
\newblock {\em CoRR}, abs/1602.03031, 2016.

\bibitem[Kin13]{primecoin_whitepaper}
Sunny King.
\newblock Primecoin: Cryptocurrency with prime number proof-of-work.
\newblock \url{https://primecoin.io/primecoin-paper.pdf}, 07 2013.

\bibitem[Nak08]{bitcoin_whitepaper}
Satoshi Nakamoto.
\newblock Bitcoin: A peer-to-peer electronic cash system.
\newblock \url{https://nakamotoinstitute.org/literature/bitcoin/}, 10 2008.

\end{thebibliography}

\end{document}